\documentclass[a4paper,11pt]{article}

\usepackage{jheppub}
\usepackage{amsmath,amssymb,amscd}
\usepackage{graphicx}
\usepackage{url}
\usepackage{nccmath} 
\usepackage{bm}

\graphicspath{ {fig/} }

\makeatletter
\def\hlinewd#1{%
  \noalign{\ifnum0=`}\fi\hrule \@height #1 \futurelet
   \reserved@a\@xhline}
\makeatother

\makeatletter
\renewcommand\@fpheader{}
\renewcommand\@journal{}
\makeatother

\definecolor{darkgreen}{rgb}{0.,.3,0}
\definecolor{darkblue}{rgb}{0.0,0.0,0.5}
\newcommand{\ep}{\epsilon}

\newcommand{\ud}{\mathrm{d}}

\title{Master Integrals for Four-Loop Massless Form Factors}

\preprint{MSUHEP-23-023, P3H-23-057, TTP23-034}

\author[a]{Roman N. Lee,}
\author[b,c]{Andreas von Manteuffel,}
\author[c]{Robert M. Schabinger,}
\author[d,e]{\\Alexander V. Smirnov,}
\author[f,e]{Vladimir A. Smirnov,}
\author[\,g]{and Matthias Steinhauser}

\affiliation[a]{Budker Institute of Nuclear Physics, 630090 Novosibirsk, Russia}
\affiliation[b]{Institut f\"ur Theoretische Physik, Universit\"at Regensburg, 93040 Regensburg, Germany}
\affiliation[c]{Department of
Physics and Astronomy, Michigan State University,\\
East Lansing, Michigan 48824, USA}
\affiliation[d]{Research Computing Center, Moscow State University,
119991, Moscow, Russia}
\affiliation[e]{Moscow Center for Fundamental and Applied Mathematics,
119992, Moscow, Russia}
\affiliation[f]{Skobeltsyn Institute of Nuclear Physics of Moscow State University,
119991, Moscow, Russia}
\affiliation[g]{Institut f{\"u}r
Theoretische Teilchenphysik,
Karlsruhe Institute of Technology (KIT),\\
76128 Karlsruhe, Germany}

\emailAdd{roman.n.lee@gmail.com}
\emailAdd{manteuffel@ur.de}
\emailAdd{schabing@msu.edu}
\emailAdd{asmirnov80@gmail.com}
\emailAdd{smirnov@theory.sinp.msu.ru}
\emailAdd{matthias.steinhauser@kit.edu}


\abstract{
We present analytical results for all master integrals for massless three-point functions, with one off-shell leg, at four loops.
Our solutions were obtained using differential equations and direct integration techniques.
We review the methods and provide additional details.
}

\begin{document}
\unitlength1cm
\maketitle
\allowdisplaybreaks[1]

\section{Introduction} 
\label{sec:intro}

Form factors are important quantities in Quantum Chromodynamics, ${\cal N}=4$ super Yang-Mills theory and other theories.
In the simplest cases, a single operator is inserted in a matrix element between two massless states, and all propagating particles are massless.
Such form factors can be constructed from vertex diagrams with two legs on the light cone, $p_1^2=p_2^2=0$, such that the corresponding Feynman integrals depend on one mass scale, 
$q^2=(p_1+p_2)^2$.
In this paper, we consider such integrals in dimensional regularization, where $d=4-2\epsilon$ is the number of space-time dimensions used to regularize ultraviolet, soft and collinear divergences.
 
 Two-loop corrections to form factors were computed more than thirty years
ago~\cite{Kramer:1986sg,Matsuura:1987wt,Matsuura:1988sm,Gehrmann:2005pd}.
The first three-loop result was presented in
Ref.~\cite{Baikov:2009bg} and later confirmed in
Ref.~\cite{Gehrmann:2010ue}. Analytic results for the three-loop
form factor integrals were presented in
Ref.~\cite{Lee:2010ik}. In Ref.~\cite{Gehrmann:2010tu}, the results of
Ref.~\cite{Lee:2010ik} were used to compute form factors at three
loops up to order $\epsilon^2$, i.e., transcendental weight eight, as
a preparation for future four-loop calculations.
These integrals and form factors have been confirmed in~\cite{vonManteuffel:2015gxa}.

Indeed, four-loop calculations have taken place since then. First analytical results for the four-loop form factors 
were obtained for the quark form factor in QCD in 
the large-$N_c$ limit, where only planar diagrams contribute~\cite{Henn:2016men}, and for the fermionic contributions~\cite{Lee:2016ixa}. All the planar master integrals for the massless four-loop form factors were
evaluated in~\cite{vonManteuffel:2019gpr}. 
The $n_f^2$ results were obtained in~\cite{Lee:2017mip}, and 
the complete contribution from color structure $(d_F^{abcd})^2$ was evaluated
in~\cite{Lee:2019zop} and confirmed in~\cite{vonManteuffel:2020vjv}. For the quark and gluon form factors, 
all corrections with three or two closed fermion loops were calculated 
in \cite{vonManteuffel:2016xki,vonManteuffel:2019wbj}, respectively, including also the singlet contributions. 
The fermionic corrections to quark and gluon form factors in four-loop QCD were evaluated in \cite{Lee:2021uqq}.
The four-loop ${\cal N}=4$ SYM Sudakov form factor
was analyzed in \cite{Boels:2017ftb} and analytically evaluated in~\cite{Lee:2021lkc}.
The complete analytical evaluation of the quark and gluon form factors in four-loop QCD was presented
in~\cite{Lee:2022nhh}. The four-loop corrections to the Higgs-bottom vertex within massless QCD were evaluated
in~\cite{Chakraborty:2022yan}.

In these calculations, two methods of evaluating master integrals were applied by our two competing groups: the method of differential equations and the evaluation by integrating over Feynman parameters: the first one was applied in
\cite{Henn:2016men,Lee:2016ixa,Lee:2017mip,Lee:2019zop} and the second one in 
\cite{vonManteuffel:2019gpr,vonManteuffel:2020vjv,vonManteuffel:2016xki,vonManteuffel:2019wbj}.  
Then our two groups combined their forces and applied these two methods when collaborating 
\cite{Lee:2021uqq,Lee:2021lkc,Lee:2022nhh,Chakraborty:2022yan}.
A crucial building block for these form factor calculations were the solutions for the four-loop master integrals, which is the topic of this paper.

\begin{figure}
\includegraphics*[width=\linewidth]{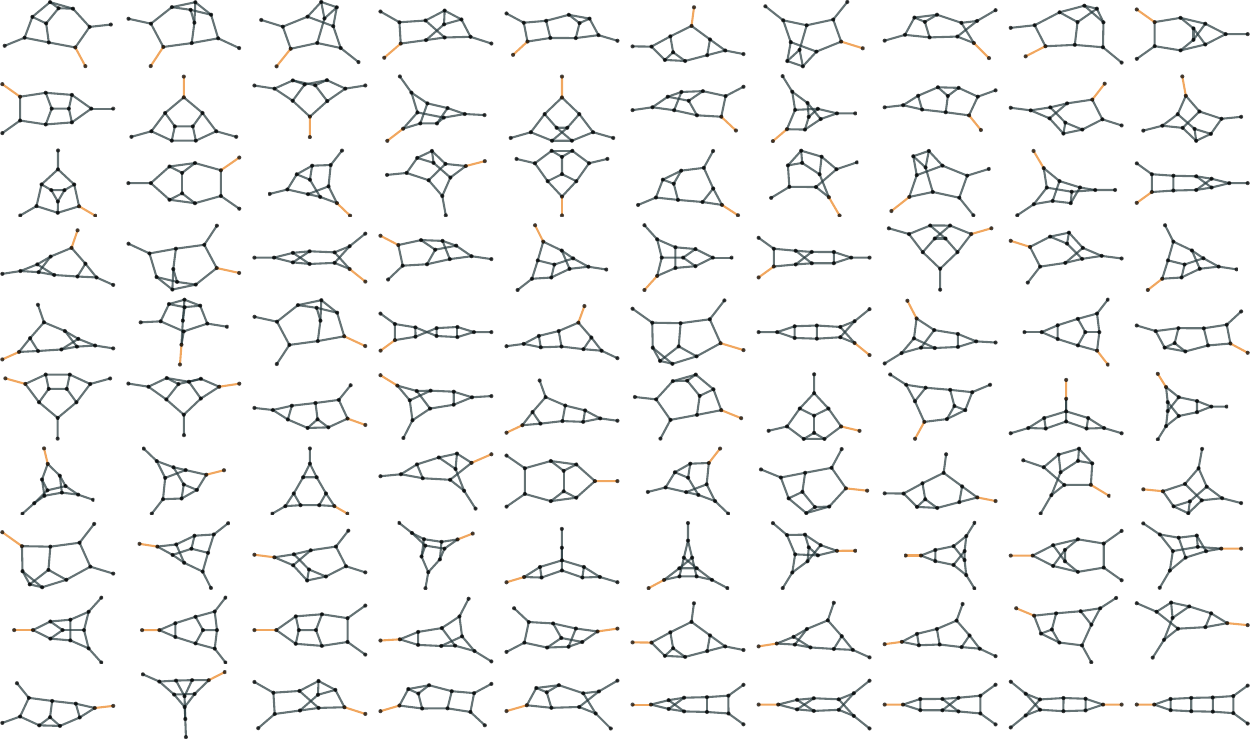}
\caption{\label{fig:alltoplevel}
Reducible and irreducible top-level topologies for four-loop form factor integrals with one off-shell leg.
}
\end{figure}

In general, four-loop form factors with one off-shell and two massless legs can involve integrals belonging to 100 reducible and irreducible top-level topologies with 12 lines, as shown in figure~\ref{fig:alltoplevel}, or sub-topologies thereof.
In this work, we present analytical solutions for the $\epsilon$ expansion of all master integrals in these topologies.
The results are given in terms of zeta values and multiple zeta values (MZV), and are complete at least up to and including weight 8, as required for N${}^4$LO calculations.

The remainder of this paper is organized as follows.
In section~\ref{sec:DE}, we describe how we applied the method of differential equations.
In a subsection, we describe peculiarities of using integration by parts (IBP) to perform reduction to master integrals.
In section~\ref{sec:FP}, we describe how we applied the method of analytical integration over Feynman parameters.
In a subsection, we discuss a dedicated reduction scheme for integrals with many dots.
In section~\ref{sec:results}, we comment on the explicit solutions for the master integrals, that we provide in the ancillary files.
In section~\ref{sec:conclusion}, we compare the two basic methods that we used.

\section{Evaluation with differential equations}
\label{sec:DE}

\subsection{Two-leg off-shell integrals, reduction to \texorpdfstring{$\epsilon$}{epsilon}-form}
The four-loop form-factor Feynman integrals that we evaluated have the following form:
\begin{equation}
\label{eq:intnorm}
    I_{\nu_1,\ldots,\nu_{18}} =  
    \frac{\Gamma^4(d/2-1)}{\pi^{2d}}
    \int \frac{1}{D_1^{\nu_1}\cdots D_{18}^{\nu_{18}}} 
   \prod_{l=1}^4 \ud^d k_l \;,
\end{equation}
where $D_{i}$ are propagators and/or numerators raised to some integer powers $\nu_i$ (indices). For the calculations presented in this section, we choose the last six indices for numerators, while the first twelve indices can be positive, i.e.\ they can correspond to propagators.
For example, for one of the most complicated diagrams for four-loop form factors
\begin{figure}[b]
\centerline{\includegraphics[scale=0.4]{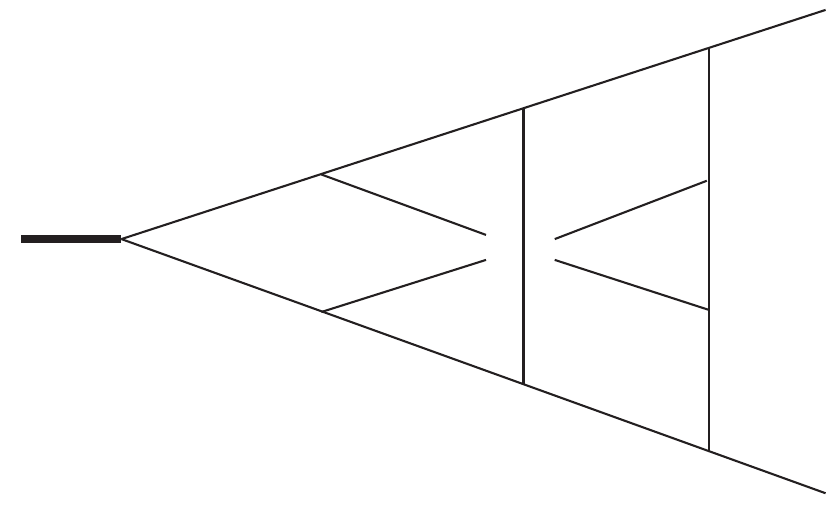}}
\caption{One of the most complicated, non-planar diagrams for four-loop form factors. This topology has four master integrals in the top-level sector.}
\label{fig:f26}
\end{figure} 
the propagators and numerators can be chosen as 
\begin{alignat}{3}
\label{eq:f26props}
D_1 &= k_1^2,&
D_2 &= (k_1 + p_1 + p_2)^2,&
D_3 &= (k_1 + k_2 - k_4 + p_2)^2,
\nonumber\\
D_4 &= (k_1 + k_2 - k_4 + p_1 + p_2)^2,\,&
D_5 &= (k_2 + k_3 + p_2)^2,&
D_6 &= (k_2 + k_3)^2,
\nonumber\\ 
D_7 &= (k_1 + k_2 + k_3 - k_4 + p_2)^2,&
D_8 &= (k_2 + k_3 - k_4 + p_2)^2,\,&
D_9 &= (k_1 + k_2 + p_1 + p_2)^2,
\nonumber\\ 
D_{10} &= k_2^2,&
D_{11} &= k_3^2,&
D_{12} &= k_4^2,
\nonumber\\ 
D_{13} &= (k_1 - k_2)^2,&
D_{14} &= (k_1 - k_3)^2,&
D_{15} &= (k_1 - k_4)^2,
\nonumber\\ 
D_{16} &= (k_2 - k_4)^2,&
D_{17} &= (k_2 + p_1)^2,&
D_{18} &= (k_3 - p_1 - p_2)^2
\end{alignat} 
where $p_1$ and $p_2$ denote the outgoing momenta of the two massless legs.

According to the strategy of IBP reduction, which was discovered more than forty years ago~\cite{Chetyrkin:1981qh,Tkachov:1981wb}, the evaluation of integrals of a given family can be reduced to the evaluation of the corresponding master integrals. In the next subsection, we describe how we did this in the case of four-loop form-factor integrals. 

Let us now turn to the method of differential equations ~\cite{Kotikov:1990kg,Gehrmann:1999as,Henn:2013pwa,Lee:2014ioa}. Since $p_1^2=p_2^2=0$, the integrals of our family depend only on one variable $q^2=(p_1+p_2)^2$, which we often set to (-1) in intermediate expressions, since this dependence is easily recovered from dimensional analysis (namely $G_{\nu_1,\ldots,\nu_{18}} \propto \left(q^2\right)^{2d-\sum \nu_i}$). In order to make use of the differential equations method, we follow the approach  of Ref. \cite{Henn:2013nsa}. We consider the family of the same topology as in figure \ref{fig:f26} now assuming that $p_2^2=xq^2$, and derive the differential system in the variable $x$. 
In what follows, we will use the terms \emph{two-scale} and \emph{one-scale} master integrals to refer to the master integrals of the family with generic $x$ and with $x=0$, respectively.
In the point $x=1$ we have $p_2^2=q^2$ and we can assume not only that $p_1^2=0$, but also that $p_1=0$, as can be clearly seen from, e.g., Feynman parametric representation. The corresponding propagator-type master integrals were evaluated in an $\epsilon$-expansion more than ten years ago~\cite{Baikov:2010hf} and are known even up to weight twelve~\cite{Lee:2011jt}. The idea is that the differential equations allow us to transfer data from the simple point $x=1$ to the desired point $x=0$.
The more involved IBP reduction of the family with $p_2^2\neq 0$ appears to be a fair price for the advantages of the differential equations method. The system of differential equations for the vector of master integrals $\boldsymbol{j}$ has the usual form
\begin{equation}
    \partial_x\boldsymbol{j}=M(\epsilon,x)\boldsymbol{j},
\end{equation} 
where $M(\epsilon,x)$ is a matrix, rational in $x$ and $\epsilon$. 

For the family in figure \ref{fig:f26} the size of the system (the number of two-scale master integrals) is as large as $374$, but even larger systems appear in other families. While not immediately obvious, a far more important characteristic of the complexity is the position of singular points in $x$. Since our final results for the one-scale master integrals involve only non-alternating MZV sums, one might speculate that the only singular points of the emerging differential systems are $x=0,\ 1,\ \infty$. And indeed, this is the case for many families that we considered. However, a few systems also contained singularities at other points. In particular, the system for the family in figure \ref{fig:f26} contained singularities for
\begin{equation}
x\in \{-1,0,1/4,1,4,\infty\}.
\end{equation}
Systems for other families contained only some of these singularities. 

Note that the singularity at $x=1/4$ is especially troublesome as it lies on the segment $[0,1]$, exactly on the integration path of the evolution operator connecting the point of interest, $x=0$, and the point $x=1$, where the boundary conditions are fixed. The \textit{general} solution of the differential system does have a branch point at $x=1/4$. From physical and technical arguments, this point can not be a branch point of the \textit{specific} solution on the first sheet (but it is a branch point on other sheets). This requirement provides yet another check of the correctness of our procedure. We may check the absence of a branch point by comparing the results obtained by shifting the integration contour slightly up and down from the real axis, which corresponds to the change $x\to x+i0$ and  $x\to x-i0$, respectively. As the coefficients of the differential system are all real, those two prescriptions are related by complex conjugation. Therefore, the absence of a branch point at $x=1/4$ can be established by checking that any of these two prescriptions leads to real-valued results. For definiteness we will assume that the integration contour is shifted up.

In order to reduce the system to $\ep$-form we use the algorithm of Refs. \cite{Lee:2014ioa,Lee2017c} (see also section 8 in Ref. \cite{Blondel:2018mad}) as implemented in \texttt{Libra} \cite{Lee:2020zfb}.
We had to introduce algebraic letters 
\begin{equation}
x_1=\sqrt{x}, \quad x_2=\sqrt{x-1/4}, \quad x_3=\sqrt{1/x-1/4}.
\end{equation}
In this way, we reduce the system to the following form:
\begin{equation}
    d\boldsymbol{J}=\epsilon\, dM\boldsymbol{J},\qquad  dM=\sum_{k=1}^{8} S_k w_k\,,\qquad  w_k=d\log P_k\,,
\end{equation} 
where
\begin{gather*}
    P_1=x,\quad P_2=1+x,\quad P_3 =1-x, \quad P_4 =4-x,\quad P_5 = 1/4-x,\\
    P_6 =1+x_1,\quad P_7=\tfrac12+ix_3,\quad P_8=\tfrac12+ix_2\,,
\end{gather*}
and $S_k$ are some constant matrices. Note that there is no variable simultaneously rationalizing $x_1,\ x_2,$ and $x_3$, as they correspond to more than $3$ square-root  branching points: ${0,\infty,4,1/4}$, see Ref. \cite{Lee2017c}. However, it appears that the weights depending on $x_1,\ x_2$ and $x_3$ never appear together in one iterated integral. More precisely, the iterated integrals which appear in our results fall into one (or a few) of the following four families:
\begin{enumerate}
    \item those containing letters in the alphabet $\{w_1,w_2,w_3\}$,
    \item those containing letters in the alphabet $\{w_1,w_3,w_6\}$,
    \item those containing letters in the alphabet $\{w_1,w_3,w_4,w_7\}$,
    \item those containing letters in the alphabet $\{w_1,w_3,w_5,w_8\}$.
\end{enumerate}
The integrals of the first two families are readily expressed via Goncharov's polylogarithms with indices $0,\pm 1$ (for the second family we have to pass to $x_1$). 
For the integrals of the third family, we pass to the variable $y_3=\frac{\sqrt3}{2x_3}$. When $x$ varies from $0$ to $1$, $y_3$ also varies from $0$ to $1$. Taking into account that 
\begin{equation}
P_1=\frac{4 y_3^2}{y_3^2+3},\quad P_3=\frac{(1-y_3)(1+y_3)}{y_3^2+3},\quad P_4=\frac{12}{y_3^2+3},\quad P_7=\frac{y_3+i\sqrt{3}}{2y_3}\,,
\end{equation}
we obtain the result for the integrals of the third family in terms of Goncharov's polylogarithms with indices $0,\pm 1,\pm i\sqrt{3}$.

\begin{figure}
    \centering
    \includegraphics[width=0.3\linewidth]{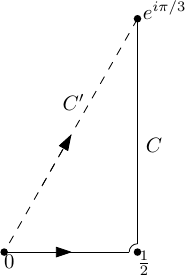}
    \caption{Integration paths $C$ and $C^\prime$ on the complex plane of $y_2$.}
    \label{fig:y2}
\end{figure}

The last family is the most complicated. We introduce the variable $y_2=P_8=\frac12+i x_2$. Taking into account our prescription $x\to x+i0$, we establish that $y_2$ follows the path $C$ depicted in figure \ref{fig:y2} when $x$ varies from $0$ to $1$.
We replace this path by the equivalent path $C^\prime$ depicted in the same figure. Since 
\begin{equation}
P_1=y_2(1-y_2),\quad P_3=y_2^2-y_2+1=(y_2-e^{i\pi/3})(y_2-e^{-i\pi/3}),\quad P_5=(y_2-1/2)^2,
\end{equation}
we obtain the result for the iterated integrals of this family in terms of Goncharov's polylogarithms with indices $0,1,e^{\pm i\pi/3}, 1/2$ and argument $e^{i\pi/3}$. We can normalize the argument to $1$ by using a homogeneity property of polylogarithms (as usual, we must exercise a certain care when dealing with polylogarithms with the trailing zeros)
\begin{equation}
    G(a_1,\ldots, a_n|a)= G(a_1/a,\ldots, a_n/a|1),\qquad a_n\neq0.
\end{equation}
Finally, the integrals of the fourth family are expressed via Goncharov's polylogarithms with indices $0,1,e^{-i\pi/3},e^{-2i\pi/3}, \frac12e^{-i\pi/3}$ and unit argument.
 
To summarize, we have the following correspondence
\begin{itemize}
    \item Families 1,2: integrals are expressed via $G(\boldsymbol{a}|1)$ with $a_k\in \{0,\pm 1\}$ (alternating MZVs),
    \item Family 3: integrals are expressed via $G(\boldsymbol{a}|1)$ with $a_k\in \{0,\pm 1,\pm i\sqrt{3}\}$,
    \item Family 4: integrals are expressed via $G(\boldsymbol{a}|1)$ with $a_k\in \{0,1,e^{-i\pi/3},e^{-2i\pi/3}, \frac12e^{-i\pi/3}\}$.
\end{itemize}
Note that the polylogarithms for the fourth family are not real-valued, so, as we explained above, the check of ``real-valuedness'' of the integrals from the fourth family provides a non-trivial check of our setup.

After we have obtained the results for the coefficients of the $\ep$-expansion of all one-scale master integrals in terms of the above mentioned polylogarithms, we used the \texttt{PSLQ} \cite{PSLQ} algorithm to recognize the results in terms of simple, non-alternating MZVs.

\subsection{IBP reduction of two-scale integrals}
\label{sec:IBP}
 
There are many public and private codes to perform IBP reduction. In this work, we applied 
the public code {\tt FIRE}~\cite{Smirnov:2014hma,Smirnov:2019qkx} and the private code {\tt Finred} by A.~von~Manteuffel.
The IBP reduction of one-scale four-loop form-factor integrals is rather complicated and the IBP reduction 
of two-scale integrals when applying the method of differential equations is even more complicated. An important point is to reveal a minimal set of master integrals. It is also important to find a basis such that the only denominators
in IBP reductions are either of the form $a d+b$, where $d$ is the space-time dimension and  $a$ and $b$ are rational numbers, or simple polynomials depending only on kinematic invariants and/or masses. Otherwise, we refer to factors in denominators as bad. To get rid of bad denominators,
i.e.\ to turn to a basis in which no bad denominators appear, one can apply the public code described 
in~\cite{Smirnov:2020quc} (see also Ref.~\cite{Usovitsch:2020jrk}).

The presence of bad denominators can essentially complicate the IBP reduction. 
It can happen that it is not possible to get rid of bad denominators. Two examples of such a situation were found in Ref.~\cite{Georgoudis:2021onj} in the context of  five-loop massless propagators.
It turned out that there was a hidden relation involving four master integrals with eleven
positive indices from four partially overlapping sectors.

For four-loop massless form factors, a similar situation takes place at the level of nine positive indices. 
This hidden relation is relevant for two
one-scale families, one of which is the family corresponding to the graph of figure~\ref{fig:f26},
\begin{equation}
\label{eq:hiddrel}
I_{0, 1, 1, 0, 1, 1, 1, 1, 1, 0, 1, 1,\ldots} =\frac{5 (d-5 ) I_{0, 1, 0, 1, 1, 0, 1, 1, 1, 1, 1, 1, \ldots} 
 + (3 d-11 ) I_{0, 1, 1, 0, 1, 0, 1, 1, 1, 1, 1, 1, \ldots}}{4 (2 d - 7)}+\ldots
 \;,
\end{equation}
where all terms from lowers sectors (of levels less than nine) are omitted and dots in the indices mean that
the last six indices are zero. The complete relation can be found in a file attached to this paper.
We have derived this relation by running FIRE with two different options (with {\tt no$\_$presolve} and without it; this option turns off partial solving of IBPs before index substitutions and thus leads to a reduction in other direction), so that it is clear that this relation is a consequence of IBP relations.
This relation has previously been depicted diagrammatically in \cite{vonManteuffel:2020vjv}, where it had been derived with {\tt Finred}, using integration-by-parts identities generated from seed integrals in a common parent topology.

Using the same strategy we have derived a hidden relation also for the two-scale integrals of this family. A file with this relation is also attached to the paper.
It has the same form as Eq.~(\ref{eq:hiddrel}) at level nine but the contribution from lower levels is different and depends on $x$. In fact, this relation should transform into the corresponding relation in the one-scale
case in the limit $x \to 0$ but to see this explicitly is more complicated in comparison to the derivation described above.
In each of the two cases, the additional relation is used to reduce the number of the master integrals. In the new basis,
all the bad denominators successfully disappear.
 
Let us mention, for completeness, that relations between a current set of the master integrals can be revealed with the help of various symmetries. This procedure is usually included into codes to solve IBP relations. An explicit example,
together with a discussion of various ways of looking for extra relations between master integrals, can be
found in~\cite{Smirnov:2013dia} in the context of the master integrals needed for the computation
of the lepton anomalous magnetic moment at three loops.
 
For the IBP reduction of one-scale integrals, both our groups applied modular arithmetic (for early discussions of such techniques see e.g.\ \cite{vonManteuffel:2014ixa,Peraro:2016wsq}).
One of our groups used the private code {\tt Finred} by A.~von~Manteuffel, which was the first code to solve IBP relations with the help of modular arithmetic and another group used {\tt FIRE}.
For the IBP reduction of two-scale integrals appearing within the method of differential equations, we applied {\tt FIRE},
also with modular arithmetic.
We first performed rational reconstruction to transition from modular arithmetic to rational numbers. Then, after fixing $d$ or $x$, we ran Thiele  reconstruction~\cite{abramowitz+stegun} to obtain a rational function of the other variable.
Since we have a good basis the denominators factor into a function of $d$ and a function of $x$.
Hence the worst possible denominator of the
coefficients of the master integrals, i.e.\ the least
common multiple of all occurring denominators, is obtained
by multiplying the univariate factors. 
Knowing the worst denominator, we could multiply the functions being reconstructed by it, and perform an iterative Newton-Newton  reconstruction~\cite{abramowitz+stegun}, i.e.\ apply two Newton reconstructions with respect to two variables.

\section{Evaluation by integration over Feynman parameters}
\label{sec:FP}

\subsection{Finite integrals, analytical integration}

Perhaps the most straightforward way to solve a Feynman integral is the direct integration of its Feynman parametric representation.
What we wish to obtain is the Laurent expansion of the integral in the regulator $\epsilon$.
We find it convenient to work with integrals which are finite for $\epsilon \to 0$, such that we can expand the \emph{integrand} in $\epsilon$ and then perform the integrations for the Laurent coefficients.
It has been shown in \cite{Panzer:2014gra,vonManteuffel:2014qoa} that one can always express an arbitrary (divergent) Feynman integral as a linear combination of a basis of ``quasi-finite'' integrals, which have convergent Feynman parameter integrations for $\epsilon \to 0$.
Requiring also the $\Gamma$ prefactor involving the superficial degree of divergence to be finite, one can also choose completely finite integrals for $\epsilon \to 0$
\cite{vonManteuffel:2015gxa}.
In this construction, the finite integrals may live in higher dimensions and may have ``dots'', i.e.\ higher powers of propagators (see \cite{Schabinger:2018dyi, Agarwal:2020dye} for generalizations of quasi-finite integrals).
A systematic list of such finite integrals can be obtained easily with the program {\tt Reduze\;2}~\cite{vonManteuffel:2012np}.
Expressing a divergent Feynman integral in terms of a basis of finite integrals, all poles in $\epsilon$ become explicit in the coefficients of this rewriting.
The explicit linear relations needed to express an integral in terms of finite basis integrals are obtained from dimension-shift identities and integration-by-parts reductions, which will be discussed in more detail below.

The integrands of the finite integrals can easily be expanded in $\epsilon$.
In general, the integrations of the coefficients can lead to complicated special mathematical functions and may be difficult to perform.
A given Feynman parametric representation for some Feynman integral can have the property of ``linear reducibility''. For linearly reducible integrals, there exists an order of integrations such that each integration can be performed in terms of multiple polylogarithms in an algorithmic way.
Thanks to the algorithms of \cite{Brown:2008um,Brown:2009ta,Panzer:2015ida} and their implementation in {\tt HyperInt}~\cite{Panzer:2014caa}, a suitable order of integration can be determined with a polynomial reduction algorithm.
If a representation is not linearly reducible, it is sometimes still possible to perform a rational transformation of the Feynman parameters such that the resulting new parametrization is linearly reducible.
For integrals resulting in elliptical polylogarithms or more complicated structures, no linearly reducible representations exists.
Currently, no algorithm is known to determine unambiguously, whether a linearly reducible parameterization exists for a given Feynman integral.

In our case, we have been able to find linearly reducible parametric representations for almost all topologies, with the only exception being two trivalent (top-level) topologies depicted as the last two entries in figure~1 of \cite{vonManteuffel:2020vjv}.
For these two topologies, the method of differential equations allows us to obtain the solutions from $\epsilon$-factorized differential equations and integrations in terms of multiple polylogarithms, as explained in section~\ref{sec:DE}.
We can not exclude that also for these topologies a linearly reducible representation could be found.
We emphasize that, in practice, direct integrations allowed us to derive complete analytical solutions through to transcendental weight 7 for \emph{all} Feynman integrals, even in those topologies which are not linearly reducible.
For the latter, this was achieved by a suitable choice of basis integrals and a high-precision numerical evaluation plus constant fitting for a single remaining integral~\cite{Agarwal:2021zft}.
The key observation was that certain Feynman integrals involve only the $\mathcal{F}$ polynomial but not the $\mathcal{U}$ polynomial at leading order in $\epsilon$, and the $\mathcal{F}$ polynomial alone could be rendered linearly reducible in all cases.

We performed the parametric integrations for the finite integrals with the {\tt Maple} program {\tt HyperInt}.
While straightforward in principle, the integration generates a large number of terms at intermediate stages.
Performance challenges arise due to bookkeeping tasks and greatest common divisor computations to combine coefficients of the same multiple polylogarithm.
In order to obtain complete information at a given transcendental weight, depending on the choice of basis integrals, one needs a different number of terms in the epsilon expansion, and each such term may require significantly different amounts of computational resources.
Usually, the first term in the $\epsilon$ expansion is relatively inexpensive to compute, and with increasing order, the computational complexity increases a lot.
For this reason, we usually start by trying out an overcomplete list of candidate integrals and compute the leading term(s) of their $\epsilon$ expansion.
We then select basis integrals, whose $\epsilon$ expansion starts with high weight and which performed well in terms of run-times for the computation of the leading term(s) in $\epsilon$.
By inserting the basis change into form factor expressions, we check for unwanted weight drops due to our choice of basis integrals.
For our basis choice, more difficult topologies start to contribute only at relatively high weight.

To perform the integrations at higher weight, in some cases, we used a parallelized {\tt HyperInt} setup on compute nodes with hundreds of GB of main memory and weeks of runtime.
In this way, we were able to analytically calculate all Feynman integrals to weight~6, all but one to weight~7 (with the last one guessed from numerical data), and a large number of integrals to weight~8.
In all cases, we checked our results with precise numerical evaluations using the program {\tt FIESTA}~\cite{Smirnov:2015mct,smirnov2021fiesta5}. The numerical evaluations were performed for our finite integrals, which allowed for better computational performance than generic integrals.

\subsection{IBP reduction of dotted integrals}

Our finite integrals typically have dots and are defined in $d=d_0-2\epsilon$ dimensions, where the reference dimension $d_0$ in many cases is larger than four, $d_0=6,8,\ldots$.
In order to express them in terms of integrals with $d_0=4$ dimensions, we exploit dimension-shift identities as described e.g.\ in \cite{Lee:2010wea}, which also introduces dotted integrals.
In particular, we employ dimension-increasing shifts which introduce four additional dots for a four-loop integral.
We establish the relation between the basis of finite integrals and a conventional basis through integration-by-parts identities, where the particular challenge lies in the reduction of the dotted integrals.

The reduction scheme is based on integration-by-parts relations in the Lee-Pomeransky representation~\cite{Lee:2013hzt,Lee:2014tja,Bitoun:2017nre}.
Defining the (twisted) Mellin transform of a function $f(x_1,\ldots,x_N)$ as
\begin{equation}
\label{eq:mellintrans}
\mathcal{M}\{f\}(\nu_1,\ldots,\nu_N)\equiv \mathcal{N} \left[ \prod_{i=1}^N \int_0^\infty \frac{x_i^{\nu_i-1}\mathrm{d}x_i}{\Gamma(\nu_i)}\right] f(x_1,\ldots,x_N),
\end{equation}
and normalizing the integral~\eqref{eq:intnorm} according to
$
I_{\nu_1,\ldots,\nu_{18}}  = 
{\mathcal{N}}\tilde{I}_{\nu_1,\ldots,\nu_{18}} /{\Gamma((L+1)d/2-\nu)}
$
one has
\begin{equation}
\tilde{I}_{\nu_1,\ldots,\nu_N} = \mathcal{M}\left\{\mathcal{G}^{-d/2}\right\}({\nu_1,\ldots,\nu_N}) ,
\quad
\text{with}
\quad
\mathcal{G}=\mathcal{U}+\mathcal{F}\,.
\end{equation}
Here, $\mathcal{U}$ and $\mathcal{F}$ are the
first and second Symanzik polynomials, respectively, $\nu=\sum \nu_i$, $N=18$, $L=4$ and $\mathcal{N}$ is an normalization constant which is not relevant in the following.

The Mellin transform~\eqref{eq:mellintrans} has the properties
\begin{align}
\mathcal{M}\{ \alpha f + \beta g \}(\bm\nu) &= \alpha \mathcal{M}\{ f \}(\bm{\nu}) + \beta \mathcal{M}\{g\}(\bm{\nu}),\\
\mathcal{M}\{ x_i f \} (\bm\nu) &=\nu_i \mathcal{M}\{f\}(\bm\nu+\bm{e}_i) ,\\
\mathcal{M}\{ -\partial_i f \} (\bm\nu) &= \mathcal{M}\{f\}(\bm\nu-\bm{e}_i) ,
\end{align}
with multi-index notation such that $\bm\nu=(\nu_1,\ldots,\nu_N)$, $\bm{e}_i=(0,\ldots,0,1,0,\ldots,0)$ has a non-zero entry at position $i$
and $\partial_i\equiv \partial/\partial x_i$.
From this it is easy to see how insertions of $x_i$ and $\partial_i$ translate to shifts of propagator powers.
We use the shift operators
\begin{align}
\left(\hat{i}^+ \tilde{I} \right)(\nu_1,\ldots,\nu_N) &= \nu_i \tilde{I}(\nu_1,\ldots,\nu_i+1,\ldots,\nu_N),\\
\left(\hat{i}^- \tilde{I} \right)(\nu_1,\ldots,\nu_N) &= \tilde{I}(\nu_1,\ldots,\nu_i-1,\ldots,\nu_N).
\end{align}
A differential operator $P$ consisting of powers of $x_i$ and $
\partial_i$ which annihilates $\mathcal{G}^{-d/2}$,
\begin{equation}\label{eq:annihilator}
P \mathcal{G}^{-d/2} = 0,
\end{equation}
generates from $\mathcal{M}\{ P \mathcal{G}^{-d/2}\} = 0$ via the substitutions
\begin{align}
x_i &\to \hat{i}^+\,,\\
\partial_i &\to - \hat{i}^-
\end{align}
a shift relation. In fact, every shift relation is generated in this way~\cite{Bitoun:2017nre}.

To construct annihilators, we make ans\"atze of the form
\begin{equation}
P = c_0 + \sum_{i=1}^N c_i \partial_i + \sum_{i,j=1}^N c_{ij} \partial_i \partial_j +\ldots\,.
\end{equation}
In the following, we will restrict ourselves to at most second order derivatives in $P$.
The functions $c_0(x_1,\ldots,x_N)$, $c_i(x_1,\ldots,x_N)$ and $c_{ij}(x_1,\ldots,x_N)$ are polynomials in the Feynman parameters $x_i$ and are determined such that \eqref{eq:annihilator} is fulfilled, which requires
\begin{align}
c_0 \left[ -\frac{2}{d}\mathcal{G}^2\right]
 + \sum_{i=1}^N c_i \left[ \mathcal{G} \partial_i \mathcal{G} \right]
 + \sum_{i,j=1}^N c_{ij} \left[ \mathcal{G} \partial_i \partial_j \mathcal{G} - \frac{(d+2)}{2} (\partial_i \mathcal{G})(\partial_j \mathcal{G})\right]
= 0\,.
\end{align}
Since the expressions in brackets are explicitly known, one can interprete this equation as a syzygy constraint for the unknown functions $c_0$, $c_i$, and $c_{ij}$.
Such syzygies can be determined algorithmically.
Once the functions $c_0$, $c_i$, and $c_{ij}$ are known, one can obtain the desired shift relations for generic $\nu_1$,\ldots,$\nu_N$ from \eqref{eq:annihilator} by replacing the Feynman parameters $x_1,\ldots,x_N$ with shift operators $\hat{1}^+,\ldots,\hat{N}^+$ in the arguments of $c_0$, $c_i$, and $c_{ij}$:
\begin{align}
\left(\left(
c_0(\hat{1}^+,\ldots)
 - \sum_{i=1}^N c_i(\hat{1}^+,\ldots) \hat{i}^-
 + \sum_{i,j=1}^N c_{ij}(\hat{1}^+,\ldots) \hat{i}^-\hat{j}^-
\right) \tilde{I} \right)(\nu_1,\ldots,\nu_N) = 0\,.
\end{align}
These equation ``templates'' are then applied to ``seed integrals'' with non-negative integer insertions for the $\nu_i$, followed by a standard ``Laporta''-style reduction of these identities for the specific loop integrals.
For the latter we use the modular arithmetic and rational reconstruction methods available in {\tt Finred}.

We compute the syzygies sector by sector, aiming at the reduction of integrals without irreducible numerators.
While we found that annihilators of linear order in the derivatives are insufficient in some cases,
annihilators of second order (involving also the $c_{ij}$) allowed us to generate the desired reductions.
Instead of computing complete syzygy modules, we restrict ourselves in the construction of the $c_0$, $c_i$, and $c_{ij}$ to a maximal degree in the Feynman parameters and employ linear algebra methods (see also \cite{Schabinger:2011dz,Agarwal:2020dye}) implemented in {\tt Finred} for their computation.

The fact that we may ignore numerators for the sector we construct the annihilator deserves a comment.
Linear annihilators produce at most a single decrementing shift operator in each term, such that no numerators will be produced for seed integrals without numerators.
This is no longer the case in the presence of a second order $(\hat{i}^-)^2$ contribution, which can indeed lead to a subsector integral with a numerator.
Interestingly, keeping also the subsector identities for a given sector, all of these auxiliary integrals can be eliminated without additional effort

In practice, we note that for subsectors with fewer lines, integrals with rather large numbers of dots need to be reduced.
On the other hand, the identities produced by the annihilator method can be reduced rather quickly.
For the present calculation, we chose to reconstruct full reduction identities with full $d$ dependence and rational numbers as coefficients, which required a large number of samples in some cases and thus non-negligible computational effort.
We found this approach attractive with regards to workflow considerations, since it decoupled our experiments with different types of basis changes from the computation of the reductions.
By storing intermediate reductions of integrals at the level of finite field samples and by reconstructing symbolic expressions only after assembling the desired linear combinations of integrals (e.g., for a specific basis change from finite field to conventional integrals), one can work with a considerably smaller number of samples and decrease the computational effort.

\section{Results in electronic files}
\label{sec:results}

We provide analytical results for the complete set of all massless, four-loop, three-point master integrals with one off-shell leg at \url{https://www.ttp.kit.edu/preprints/2023/ttp23-034/}\,.
Please see the file {\tt README} for details regarding the employed conventions and for a description of the various files.

Our analytical results for the vertex integrals with one off-shell leg are given as Laurent expansions in $\epsilon$ and allow quantities to be computed at least up to and including weight~8, in many cases up to and including weight~9.
Results obtained by the method of differential equations are given in a UT basis (strictly speaking, the UT property is a conjecture at higher orders in $\epsilon$).
The complete set of all master integrals is given in terms of finite integrals, which allows for easier numerical checks.
In addition, we provide mappings to a more conventional ``Laporta basis'', determined by a generic ordering of integrals.

We also provide results for the vertex integrals with two off-shell legs, which we used to employ the method of differential equations.
In particular, we define basis integrals, which lead to $\epsilon$-factorized differental equations.
Moreover, we also provide the differential equations themselves.

For the calculation of some (physical) quantity to weight 8 using some non-UT basis, it may seem that one needs information from higher order terms in the $\epsilon$ expansion that are not provided here.
To work around this problem, we introduced tags in the expansions representing specific unknown higher weight contributions.
By expanding to sufficiently high order in $\epsilon$ and making sure that these tags drop out in the final result, one can still use such an alternative basis.

\section{Conclusion}
\label{sec:conclusion}

We presented solutions for all four-loop master integrals contributing to massless vertex functions with one off-shell leg.
Our results for the Laurent expansion in the dimensional regulator $\epsilon$ are given in terms of regular and multiple zeta values and are complete up to and including at least transcendental weight eight.
We provide concise definitions of all master integrals, their analytical solutions, basis transformations and further auxiliary expressions in electronic files, that can be downloaded from
\url{https://www.ttp.kit.edu/preprints/2023/ttp23-034/}\,.

We employed two methods to obtain these results: one based on differential equations for topologies with an additional off-shell leg, one based on direct parametric integrations of finite integrals.
In a large number of cases, we employed both methods to compute integrals in the same topology.
Moreover, almost all integrals were checked up to weight seven by such redundant calculations.
For the weight six and weight seven contributions we also had non-trivial checks available from the cusp and collinear anomalous dimensions extracted from the poles of different form factors.
Various weight eight contributions have been obtained using only one of the two described methods, those we checked against precise numerical evaluations with {\tt FIESTA}.

We observed that it can be computationally rather inexpensive to compute lower weight contributions in the parametric integration approach, once one has a linearly reducible representation available.
Furthermore, a suitable basis choice can avoid contributions from more complicated topologies at lower weight.
For that reason, the weight seven contributions could essentially be obtained from direct integrations.
At weight eight, the situation is different.
For two particularly complicated top-level topologies we could not even find a suitable starting point, that is, a linearly reducible representation.
Moreover, for a number of other topologies, we were not successful with direct integrations due to the high computational resource demands.

Remarkably, in all these challenging cases, the differential equation approach worked well and allowed us to obtain analytical solutions.
Despite the fact, that one generalizes the problem by taking one of the light-like legs off-shell, the power of the method outweighed this potential drawback in practice.
As a bonus, the method allowed us to arrive at uniformly transcendental basis integrals, and one can obtain results also at even higher transcendental weight if needed.

\section*{Acknowledgments}

AvM and RMS gratefully acknowledge Erik Panzer for related collaborations.
This research was supported by the Deutsche Forschungsgemeinschaft (DFG,
German Research Foundation) under Grant No.\ 396021762 — TRR 257 ``Particle Physics
Phenomenology after the Higgs Discovery'' and by the National Science Foundation (NSF) under Grant No.\ 2013859 ``Multi-loop amplitudes and precise predictions for the LHC''. The work of AVS and VAS was supported by the Russian Science Foundation under Agreement No. 21-71-30003
(IBP reduction) and by the Ministry of Education and Science of
the Russian Federation as part of the program of the Moscow Center for Fundamental and Applied Mathematics under Agreement No. 075-15-2022-284
(numerical checks of results for the master integrals with {\tt FIESTA}).
We acknowledge the High Performance Computing Center at Michigan State University
for computing resources.
The work of RNL was supported by the Russian Science Foundation, grant No.  20-12-00205.
The diagram in figure~\ref{fig:f26} was drawn with the help of {\tt Axodraw}~\cite{Vermaseren:1994je}
and {\tt JaxoDraw}~\cite{Binosi:2003yf}.

\bibliographystyle{JHEP}
\bibliography{ff4lMI}
 
\end{document}